\documentclass[journal]{IEEEtran}

\usepackage[T1]{fontenc}
\usepackage{amssymb}
\usepackage{amsmath}
\usepackage{algorithm}
\usepackage{algorithmic}
\usepackage{graphicx}
\usepackage{epstopdf}
\DeclareGraphicsExtensions{.eps}
\usepackage{caption}

\usepackage{subfigure}
\usepackage{cite}

\renewcommand\IEEEkeywordsname{Index Terms}

\begin{document}
%
\title{Max-Min Fair Hybrid Precoding for Multi-group Multicasting in Millimeter-Wave
Channel}
\author{Fawwaz Alsubaie
}

\maketitle

\begin{abstract}
The potential of using millimeter-wave (mmWave) to encounter the current bandwidth shortage has motivated packing more antenna elements in the same physical size which permits the advent of massive multiple-input-multiple-output
(MIMO) for mmWave communication. However, with increasing number of antenna elements, the ability of allocating a single RF-chain per antenna becomes infeasible and unaffordable. As a cost-effective alternative, the design of hybrid precoding has been considered where the limited-scattering signals are captured by a high-dimensional RF precoder realized by an analog phase-shifter network followed by a low-dimensional digital precoder at baseband. In this paper, the max-min fair problem is considered to design a low-complexity hybrid precoder for multi-group multicasting systems in mmWave channels. The problem is non-trivial due to two main reasons: the original max-min problem for multi-group multicasting for a fully-digital precoder is non-convex, and the analog precoder places constant modules constraint which restricts the feasible set of the precoders in the design problem. Therefore, we consider a low complexity hybrid precoder design to tackle and benefit from the mmWave channel structure. Each analog beamformer was designed to maximize the minimum matching component for users within a given group. Once obtained, the digital precoder was attained by solving the max-min problem of the equivalent channel.
\end{abstract}


\IEEEpeerreviewmaketitle

\section{INTRODUCTION}

The event of common data targeting a mass of users or relay stations, in the case of relay networks, has become popular in current wireless systems. Services like video/audio streaming, news, common clips and messages are expected to further grow in next-generation wireless systems \cite{Silva2009}. In a scenario where a group of users demand the same information, i.e. single-group multicasting (SGM) or broadcasting,
the base station (BS) can provide the service by a point-to-multipoint connection. In the case of multi-group multicasting (MGM), groups with disjoint sets of users are entitled to different common messages. In the case of a single user per group, the scenario is known as multi-user (MU) downlink beamforming. The provision of multicasting services,
both SGM and MGM, has been introduced by the Global System for Mobile communication (GSM) and Universal Mobile Telecommunications System (UMTS) as a form of multimedia broadcast/multicast service \cite{Ogunbekun2003,Bakhuizen2005}.
In addition, a form of multicasting has been proposed by the Worldwide Interoperability for Microwave Access (WiMAX) \cite{Jiang2007}. 

When considering a serving-network where users can request same common messages (multicasting), the support of such services can be provided under a wired network. However, from a cost and feasibility prospective, it is impossible to extend cables to every user or a relay station. The solution is to utilize the wireless medium to perform multicasting. The apparent wireless nature as a broadcast medium causes cross-talk or co-channel interference among users scheduled at the same time and frequency slots, not mentioning fading and shadowing. Then, to
mitigate interference, users used to be scheduled among the orthogonal resources, e.g. time and frequency. This resulted in increasing network traffic and decreasing spectral efficiency. 

To address this issue, in MU-beamforming, smart antenna array with adjustable antenna weights has been implemented at the transmitter to beamform, or precode, to the served users. Hence, users can be simultaneously scheduled at the same time-frequency slot while experiencing minimal interference. The later technique utilizes the third degree of freedom (DoF) in the system, i.e. space, and formally called spatial
division multiple access (SDMA). Techniques to perform SDMA to design optimal and sub-optimal precoders and with various scenarios have been extensively studied in the literature \cite{Bengtsson1999,MBengtsson2002,Visotsky1999,Rashid-Farrokhi1998,Yu2007,Schubert2004}. When considering multicasting, the concept of physical-layer multicasting (PHY-multicasting) was firstly introduced in \cite{Sidiropoulos2006} for SGM and in \cite{Karipidis2008} for MGM. Moreover, the discussed design problems were far from trivial compared to MU-beamforming and
are detailed later in this paper.

In addition to the future need for multimedia/ multicasting services, the demand for low-latency application and delivery of high-quality multimedia content is a genuine challenge. With the rapid growth of smartphones, available carrier frequency slots have become ever more scarce. To overcome the bandwidth shortage, the underutilized millimeter wave (mmWave) has been strongly suggested for the next-generation frequency spectrum use \cite{Lu2014}. The mmWave signal experiences orders-of-magnitude path loss due to a ten-fold carrier frequency increase. With this difficulty comes the interesting feature of mmWave that is packing more antennas at the same physical size of the original microwave antenna. Large antenna array offers beamforming gain to overcome path loss and allows spatial multiplexing which could improve the overall system spectral efficiency \cite{Heath2016,Rappaport2013,Doan2004}. 

So far, the adaptive antenna array permits flexible configuration in the RF domain of both magnitude and phase. However, in mmWave channels, the BS is equipped with a large antenna array and it is rather infeasible to dedicate a single RF-chain per antenna element \cite{Pi2011}. Therefore, precoding has been divided into two stages: digital low-dimensional baseband precoding followed by high-dimensional RF-precoding realized by an analog phase shifter network. Since the introduction of hybrid precoding in \cite{Zhang2005}, it has been a strong candidate for the next-generation wireless system \cite{Alkhateeb2014}. In the hybrid structure, the RF-precoding has a fewer number of RF-chains fully or partially-connected with all antennas whereas the digital-processing is done at the baseband. The Full analysis of the minimal number of RF chains to realize a fully-digital precoder is given in \cite{Zhang2014,Bogale2016,Chae2015}. Works on designing a hybrid precoder in mmWave channels for various
systems are given in \cite{Ayach2014,Ayach2012,Alkhateeb2015,Sohrabi2015,Alkhateeb2013,Yu2015} while a low-complexity hybrid precoder design under independent and identically distributed (i.i.d) Rayleigh fading channel is considered in \cite{Liang2014}.

\section{Related Previous Work }

In multicasting, SGM or MGM, the existing work have addressed various
frameworks to specify what-called the optimal precoder. Herein, we
mention the existing design problems for multicasting systems:
\begin{itemize}
\item Quality of Service (QoS): the QoS problem considers a minimum service
level for each individual user in the multicasting system. The goal
is to minimize the total transmit power subject to the minimum service
levels for all users in the multicasting system. The optimal precoder
in the QoS sense provides the minimum service level(s) with the minimum
power consumption. 
\item Max-min fair (max-min): the max-min fair problem is concerned with
a fair performance in a multicasting system. The goal is to maximizes
the minimum QoS level(s) for all users in the multicasting system
subject to a total transmit power. Conceptually, this is achieved
by reducing the power for the good-condition-channel users compared
to the worst channel(s). The optimal precoder in the fairness sense
ensures a fair performance among users and satisfies the total transmit
power constraint with equality.
\item Sum-Rate Maximization (SR): the SR problem considers the optimal precoder
for which the multicasting system experiences the ultimate throughput.
Along this line, the SR problem doesn't consider fairness among users
or groups. For example, in the case of MGM, low-channel-condition
group can be set to service unavailability. In other words, the power
is not consumed to compensate for channel conditions.
\end{itemize}
One or more of the above problems could be solved under Per Antenna
Constraint (PAC). The motivation of investigating such a constraint
comes from a practical system implementation aspect. Power flexibility
is not always feasible at the transmitter due to different antennas
having individual amplifiers and hence the need to specify the power
consumed by each antenna element. Also, different antenna amplifiers
can have different power range, i.e. different saturation power levels,
and thus PAC problem can help in controlling the power of each antenna
element.

\subsection{Single-group multicasting (SGM):}

In SGM, the cell performance characterized by total transmission power
or overall throughput is constrained by the worst-condition user in
the group. In SGM, a single common message addresses a group of co-channel
users and thus cross-talk is not an issue. While SGM can be a special
case of MGM and a work on MGM would generally apply on SGM, we mention
specific works on SGM systems. 

Physical layer multicasting was firstly introduced by Lopez in \cite{Lopez2002},
where the sum of signal-to-noise ratio (SNR) was maximized for all
users. The optimization was equivalent to maximizing an average SNR
not considering individual users which boiled down to a simple eigenvalue
problem. However, the work in \cite{Lopez2002} has inspired the development
of various algorithms and design problems concerning multicasting
systems.

The QoS problem for SGM has been firslty introduced in \cite{Sidiropoulos2006},
where the problem is formulated as a non-convex Quadratically Constrained
Quadratic Program (QCQP) and shown to be an NP-hard problem. The problem
is relaxed, new variables are introduced, and reformulated as a semi-definite
relxation program (SDR) \cite{Luo2010}. Following that, if the solution
is not rank-one, Gaussian randomization with scaling are used to find
the optimal precoder. The max-min problem has also been addressed
in \cite{Sidiropoulos2006}, and the solution is found following the
previous strategy.

In \cite{Dai2016}, the max-min fair problem has been formulated to
find a sub-optimal hybrid precoder design under mmWave channel. The
RF-precoder and baseband precoder are decoupled. The RF-precoder is
designed based on a codebook to maximize an upper bound assuming a
fixed baseband precoder. Once obtained, the digital precoder is obtained
following the strategy in \cite{Sidiropoulos2006}.

\subsection{Multi-group multicasting (MGM):}

A fundamental difference in MGM compared to SGM: is the existing of
disjoint sets of users demanding different common messages and hence
interference becomes an issue. However, the problem is very general
where it includes many scenarios, SGM, for example. In addition, the
SGM design problems are always feasible, whereas for MGM systems,
the design problems can be infeasible. On the other hand, some of
the MGM design problems are more flexible in which the service levels
for different groups can be adjusted to fulfill some optimization
goals.

The QoS and MMF problems for MGM have been firstly introduced in \cite{Karipidis2008}.
Due to interference and possible infeasibility, the QoS problem for
MGM is different from the SGM one presented in \cite{Sidiropoulos2006}.
The SDR method is proposed to solve the QoS problem similar to \cite{Sidiropoulos2006}.
However, the randomization is more involved since the scaling can
enhance the interference. Therefore, a power control program is solved
to provide the right scaling factors. For the max-min problem, the
SDR program is non-linear and is solved, if feasible, by a bisection
method followed by a multi-group power control program. In the case
of Vandermonde channels, which is the case in line-of-sight (LoS)
scenarios, the channel matrices are shown to be rank-one and the relaxation
is tight. Hence, if the problem is feasible, the optimal solution
is always obtainable \cite{Karipidis2007}. 

To account for practical system limitation, the MGM sum-rate problem
with PAC has been presented in \cite{Christopoulos2014a} while the
max-min fair problem with PAC has been discussed in \cite{Christopoulos2014b,Christopoulos2014}.

In order to mitigate SDR complexity, specially as number of users
increases, Successive Convex Approximation (SCA) has been proposed to solve the QoS problem in \cite{Universit2011}
and max-min fair problem in \cite{Schad2012}. Also, the max-min sum
rate and max-min with PAC have been investigated in \cite{Kaliszan2012}
and \cite{Christopoulos}, respectively. Similar to SGM, the solution
might not be optimal but it is a low-complexity design compared to SDR.

Recent works have considered hybrid precoding for MGM systems under mmWave channels \cite{Mx1,Mx2,QoS1,QoS2}.

In this paper, we consider the problem for MGM
systems with hybrid precoding introduced in \cite{Zhang2005} under
mmWave channels. We give a low complexity max-min fair hybrid
precoding design for MGM systems under mmWave channels. System performance
is simulated and discussed. 

\emph{Notations:} We use the following notation throughout this paper: $\mathbf{A}$
is a matrix; $\mathbf{a}$ is vector; $a$ is a scalar; $\mathbf{I}_{N}$
is the $N\times N$ identity matrix; $\parallel\mathbf{.}\parallel_{{\scriptscriptstyle F}}$
is the Frobenius norm; $(\mathbf{.})^{\textrm{T}}$ is the transpose
operator; $(\mathbf{.})^{H}$ is the conjugate transpose operator;
$\mathrm{tr}(\mathbf{.})$ is the trace operator; $\mathbb{E}\left[\mathbf{.}\right]$
is the expectation and $\arg(\mathbf{.})$ is the phase of the entries
for matrix or vectors. 

\section{System Model}

Consider a single-cell system where the base station (BS) is equipped
with $N$ antenna elements and $N_{\mathrm{{\scriptscriptstyle RF}}}$
RF-chains communicating with $M$ single-antenna users. Further, let
there be a total of $1\leq G\leq N_{{\scriptscriptstyle \mathrm{{\scriptscriptstyle RF}}}}$
multicasting groups where the multicast group index set writes as
$\mathcal{I}=\left\{ \mathcal{G}_{1},\ldots,\mathcal{G}_{G}\right\} $
and $\mathcal{G}_{k}$ is the set of users belonging to the $k$th
multicast group, $k=\left\{ 1,\ldots,G\right\} $. Each user can belong
to one group only and hence, $\mathcal{G}_{i}\bigcap\mathcal{G}_{j}=0$,
$\forall i,j\in\left\{ 1,\ldots,G\right\} $ and $\bigcup_{i=1}^{G}\mathcal{G}_{i}=M$.
We assume a single RF chain is dedicated to each group and there are
as many groups as number of chains at a given time instant and thus
$N_{\mathrm{{\scriptscriptstyle RF}}}=G$. 

On the downlink, the base station applies the $N_{{\scriptstyle \mathrm{{\scriptscriptstyle RF}}}}\times G$
digital baseband precoder $\mathbf{W}=[\mathbf{w}_{1},\mathbf{w}_{2},\ldots,\mathbf{w}_{G}]$
to the sampled transmitted data $\mathbf{s}\in\mathbb{\mathbb{C}}^{G\times1}$
and up-converts the processed signal to the carrier frequency by applying
the $N\times N_{{\scriptstyle \mathrm{{\scriptscriptstyle RF}}}}$
RF precoder $\mathbf{F}=[\mathbf{f}_{1},\mathbf{f}_{2},\ldots,\mathbf{f}_{N_{{\scriptscriptstyle {\scriptscriptstyle \mathrm{RF}}}}}]$.
The precoded transmitted signal $\mathbf{x}\in\mathbb{C}^{N\times1}$
writes as:
\begin{equation}
\mathbf{x}=\mathbf{FWs},\label{eq:1}
\end{equation}
where $\mathbb{E}\left[\mathbf{s}\mathbf{s}^{\ast}\right]=\mathbf{I}_{G}$.
The RF precoder $\mathbf{F}$ is implemented using an analog phase
shifter network and its entries has the constant modulus constraint
such that $\mathbf{F}(i,j)=e^{\theta_{i,j}}$ and $\mid\mathbf{F}(i,j)\mid^{2}=1$.
The total transmission power $P$ is allocated such that $\parallel\mathbf{FW}\parallel_{{\scriptscriptstyle F}}^{2}=P$.

For simplicity, we focus on the narrow-band block fading channel model
in which the $m$th user, belonging to the $k$th group, observes
the following received signal:
\begin{equation}
y_{m}=\underbrace{\mathbf{h}_{m}^{H}\mathbf{F}\mathbf{w}_{k}s_{k}}_{\textrm{Intended-signal}}+\underbrace{\underset{i\neq k,i=1}{\overset{G}{\sum}}\mathbf{h}_{m}^{H}\mathbf{F}\mathbf{w}_{i}s_{i}}_{\textrm{Interference-term}}+\underbrace{n_{m}}_{\textrm{noise-term}},\label{eq:2}
\end{equation}
where $\mathbf{h}_{m}\in\mathbb{C}^{N\times1}$ is the mmWave channel
response vector between the BS antenna elements and the $m$th single-antenna
user. We adopt a geometric finite-scattering channel model with $L$
propagation paths between the BS and a mobile terminal. The mmWave
channels are expected to have finite scattering for a single propagation
path and hence we assume that each scatter contribute a single path
between the BS and a mobile user. Additionally, we consider an Uniformally-Linear-Array
(ULA) structure at the BS. Then, the channel for the $m$th user is
written as:
\begin{equation}
\mathbf{h}_{m}=\sqrt{\frac{N}{L}}\sum_{l=1}^{L}\beta_{m,l}\mathbf{a}_{t}(\phi_{m,l}),\label{eq:3}
\end{equation}
where $\mathbf{a}_{t}(\phi_{m,l})\in\mathbb{C}^{N\times1}$ is the
array response vector at the base station due to the transmitted (beamformed)
signal at the $\phi_{m,l}$ angle of departure (AoD) and can be expressed
as:
\[
\mathbf{a}_{t}(\phi_{m,l})=\frac{1}{\sqrt{N}}\left[1,e^{\textrm{j}2\pi\frac{d}{\lambda}\sin(\phi_{m,l})},\ldots,e^{\textrm{j}(N-1)2\pi\frac{d}{\lambda}\sin(\phi_{m,l})}\right]^{\textrm{T}},
\]
and $\beta_{m,l}$ is the channel gain of the $m$th user in the direction
of the steered signal. The term $n_{m}$ is an Additive White Gaussian
Noise (AWGN) at the $m$th receiver such that $\mathbb{E}\left[n_{m}n_{m}^{\ast}\right]=\sigma_{m}^{2}$,
and we assume without the lose of generality $\sigma_{1}^{2}=\sigma_{2}^{2}=\cdots=\sigma_{M}^{2}=\sigma_{n}^{2}$.
The Signal to Interference plus Noise Ratio (SINR) experienced by
the $m$th is given by:
\begin{equation}
\mathrm{SINR}{}_{m}=\frac{\left|\mathbf{h}_{m}^{H}\mathbf{F}\mathbf{w}_{k}\right|^{2}}{\left|\underset{i\neq k,i=1}{\overset{G}{\sum}}\mathbf{h}_{m}^{H}\mathbf{F}\mathbf{w}_{i}\right|^{2}+\sigma_{n}^{2}}\label{eq:SINR}
\end{equation}
Assuming i.i.d Gaussian input-streams, the instantaneous rate experienced
by the $m$th user writes as:
\begin{equation}
R_{m}=\log_{2}(1+\mathrm{SINR}_{m})\label{eq:rate}
\end{equation}

\section{PROBLEM FORMULATION} \label{system}

The main objective is to design a low-complexity hybrid precoder for
multi-group multicasting systems in mmWave channels. In this work,
we consider the max-min fairness problem to design the hybrid precoder
to ensure minimum service level for all users regardless to the group
they belong to. The max-min fair problem is formally defined as:
\begin{eqnarray}
 &  & \underset{{\scriptscriptstyle \mathbf{F},\mathbf{W}}}{\max}\min_{k\in\{1,\ldots,G\}}\min_{m\in\mathcal{G}_{k}}\quad\frac{\left|\mathbf{h}_{m}^{H}\mathbf{F}\mathbf{w}_{k}\right|^{2}}{\left|\underset{i\neq k}{\overset{}{\sum}}\mathbf{h}_{m}^{H}\mathbf{F}\mathbf{w}_{i}\right|^{2}+\sigma_{n}^{2}}\label{eq:p1}\\
\mathrm{s.t.} & : & \mathbf{F}\in\mathcal{F},\nonumber \\
 &  & \left\Vert \mathbf{FW}\right\Vert _{{\scriptscriptstyle F}}^{2}\leq P,\nonumber 
\end{eqnarray}
where $\mathbf{\mathcal{F}}$ is the set of $N\times G$ matrices
with constant modules entries and $\mathbf{W}$ is the $N_{\mathrm{{\scriptscriptstyle RF}}}\times G$
digital precoder. The total transmission power constraint is explicitly
specified in (\ref{eq:p1}). In the fully digital system, i.e, when
$\mathbf{FW}=\mathbf{V}_{\mathrm{D}}$ where $\mathbf{V}_{\mathrm{D}}\in\mathbb{C}^{N\times G}$,
the problem in (\ref{eq:p1}) has been shown to be non-convex (NP-hard)
and thus cannot be solved efficiently using convex optimization solvers,
e.g, interior-point method \cite{Karipidis2008}. Moreover, restricting
the search space by adding the special structure of the RF precoder
results in an even more difficult problem. In addition, the baseband
precoder $\mathbf{W}$ needs to be jointly designed with the RF precoder
$\mathbf{F}$ and the optimization is often found intractable \cite{Ayach2014}.
Therefore, it is often advised to decouple the problem into two parts.
First, we design the RF precoder $\mathbf{F}$ by assuming a fixed
baseband precoder $\mathbf{W}$ and then $\mathbf{F}$ is fixed and
$\mathbf{W}$ is designed \cite{Yu2015}. Assuming $\mathbf{F}^{\star}$
is attained, and for clarity we denote it as $\mathbf{F}$, the program
in (\ref{eq:p1}) can be re-written as:
\begin{eqnarray}
\mathbf{W^{\star}} & = & \underset{{\scriptscriptstyle \mathbf{W}}}{\arg\max}\min_{k\in\{1,\ldots,G\}}\min_{m\in\mathcal{G}_{k}}\quad\frac{\left|\mathbf{h}_{\mathrm{\mathbf{eff}},m}^{H}\mathbf{w}_{k}\right|^{2}}{\left|\underset{i\neq k}{\overset{}{\sum}}\mathbf{h}_{\mathrm{\mathbf{eff}},m}^{H}\mathbf{w}_{i}\right|^{2}+\sigma_{n}^{2}} \nonumber\\
\mathrm{s.t.} & : & \mathbf{W}\in\mathbb{C}^{N_{\mathrm{RF}}\times G},\label{eq:p2} \\
 &  & \left\Vert \mathbf{FW}\right\Vert _{{\scriptscriptstyle F}}^{2}\leq P,\nonumber 
\end{eqnarray}
where $\mathbf{h}_{\mathbf{eff},m}^{H}=\mathbf{h}_{m}^{H}\mathbf{F}$.
The program in (\ref{eq:p2}) can be equivalently written as a Semi-Definite
Program (SDP). Defining $\left\{ \mathbf{Q}_{\mathbf{eff},m}=\mathbf{h}_{\mathbf{eff},m}\mathbf{h}_{\mathbf{eff},m}^{H}\right\} _{m=1}^{m=M}$
and $\left\{ \mathbf{X}_{k}=\mathbf{w}_{k}\mathbf{w}_{k}^{H}\right\} _{k=1}^{k=G}$
. Thus, $\left|\mathbf{h}_{\mathbf{eff},m}^{H}\mathbf{w}_{k}\right|^{2}=\mathrm{tr}(\mathbf{Q}_{\mathbf{eff},m}\mathbf{X}_{k})$,
$\mathbf{X}_{k}\geq\mathbf{0}$ and noting $\left\Vert \mathbf{FW}\right\Vert _{{\scriptscriptstyle F}}^{2}=\sum_{k=1}^{G}\mathrm{tr}(\mathbf{F}\mathbf{X}_{k}\mathbf{F}^{H})$,
(\ref{eq:p2}) is then reformulated as:
\begin{eqnarray}
\left\{ \mathbf{X}_{k}^{\star}\right\} _{k=1}^{G} & = & \underset{\left\{ \mathbf{X}_{k}\right\} _{k=1}^{G}}{\arg\max\quad t}\label{eq:p3}\\
\mathrm{s.t.} & : & \frac{\mathrm{tr}(\mathbf{Q}_{\mathbf{eff},m}\mathbf{X}_{k})}{\sum_{i\neq k}\mathrm{tr}(\mathbf{Q}_{\mathbf{eff},m}\mathbf{X}_{i})+\sigma_{n}^{2}}\geq t,\nonumber \\
 &  & \sum_{k=1}^{G}\mathrm{tr}(\mathbf{F}\mathbf{X}_{k}\mathbf{F}^{H})\leq P,\nonumber \\
 &  & \mathbf{X}_{k}\geq\mathbf{0},t\geq0,\nonumber \\
 &  & \textrm{rank}(\mathbf{X}_{k})=1,\nonumber \\
 &  & \forall k\in\left\{ 1,\ldots,G\right\} ,\forall m\in\left\{ 1,\ldots,M\right\} .\nonumber 
\end{eqnarray}
where (\ref{eq:p3}) is a reminiscent of the max-min fairness problem
discussed in \cite{Karipidis2008} where the non-convex rank-one constraints
are dropped and the problem is solved, if feasible, by bisection to
yield an upper bound for the maximum SINR experienced by all users.
The relaxed problem is re-casted as:
\begin{eqnarray*}
\mathcal{P}_{main} & : & \underset{\left\{ \mathbf{X}_{k}\right\} _{k=1}^{G}}{\mathrm{\max}\quad t}\\
\mathrm{s.t.} & : & \mathrm{tr}(\mathbf{Q}_{\mathbf{eff},m}\mathbf{X}_{k})-t\left(\sum_{i\neq k}\mathrm{tr}(\mathbf{Q}_{eff,m}\mathbf{X}_{i})+\sigma_{n}^{2}\right)\geq0,\\
 &  & \sum_{k=1}^{G}\mathrm{tr}(\mathbf{F}\mathbf{X}_{k}\mathbf{F}^{H})\leq P,\\
 &  & \mathbf{X}_{k}\geq0,t\geq0,\\
 &  & \forall k\in\left\{ 1,\ldots,G\right\} ,\forall m\in\left\{ 1,\ldots,M\right\} .
\end{eqnarray*}

Therefore, in this paper, we aim to design a low-complexity hybrid
precoder design for multi-group multicasting systems to approach the
upper bound achieved by the fully-digital precoder bearing in mind
the special structure of the channel, namely, sparse multipath channels
or mmWave channels. 

\section{PROPOSED SOLUTIONS} \label{HP}

Problem $\mathcal{P}_{main}$ can be solved to attain a local optimal
digital baseband precoder $\mathbf{W}^{\star}$. However, prior solving
for $\mathbf{W}^{\star}$, we need to design the RF precoder and construct
the effective channel(s) $\mathbf{Q}_{\mathbf{eff}}$. Fortunately,
the nature of the mmWave suggests a way in selecting the RF precoder
and here we could indicate the following remarks: 
\begin{itemize}
\item In the case of a mmWave channel between the BS and a served user
and as number of transmit antenna increases, the dependence on other
paths becomes less important compared to the strongest path \cite{Alkhateeb2015}.
Thus, we consider single path mmWave channels.
\item In single path mmWave channels and as number of transmit antenna
increases, different users with distinct AoD's from the BS exhibit
orthogonal channel vectors, thanks to the asymptotic orthogonality
property of mmWave channels \cite{Ayach2012}.
\item In order to realize a fully-digital precoder under any channel structure,
\cite{Sohrabi2015,Zhang2005} have shown that two RF-chains are needed
per a digital precoder, i.e, $G=2N_{{\scriptscriptstyle \mathrm{RF}}}$.
In addition, in order to modulate $G$ streams, it is necessary to
have at least $G$ RF-chains and hence $N_{\mathrm{{\scriptscriptstyle RF,min}}}=G$.
Therefore, we consider a single RF-chain per group, $N_{\mathrm{{\scriptscriptstyle RF}}}=G$. 
\item In single path mmWave channels of users with distinct angles from
the BS and as $N\rightarrow\infty$, the conjugate analog beamformer
per user, $N_{\mathrm{{\scriptscriptstyle RF}}}=M$, is optimal \cite{Liang2014a}. 
\item In \emph{connection} of the last mentioned remarks, we propose to
design the RF precoder as in the following program:
\end{itemize}
\begin{eqnarray}
 &  & \underset{\mathbf{f}_{k}}{\mathrm{\max}}\min_{m\in\mathcal{G}_{k}}\quad\left|\mathbf{h}_{m}^{H}\mathbf{f}_{k}\right|^{2}\label{eq:f1}\\
\mathrm{s.t.} & : & \left|\mathbf{f}_{k}(i)\right|=1,\nonumber \\
 &  & \forall k\in\left\{ 1,\ldots,G\right\} ,\forall i\in\left\{ 1,\ldots,M\right\} .\nonumber 
\end{eqnarray}

However, the program in (\ref{eq:f1}) is NP-hard due to the non-convex
constant modules constraint. Therefore, we opt to relax the constraint
as shown in (\ref{eq:f2}).

\begin{eqnarray}
 &  & \underset{\mathbf{u}_{k}}{\mathrm{\max}}\min_{m\in\mathcal{G}_{k}}\quad\left|\mathbf{h}_{m}^{H}\mathbf{u}_{k}\right|^{2}\label{eq:f2}\\
\mathrm{s.t.} & : & \left|\mathbf{u}_{k}\right|^{2}=N,\nonumber \\
 &  & \forall k\in\left\{ 1,\ldots,G\right\} ,\forall m\in\left\{ 1,\ldots,M\right\} .\nonumber 
\end{eqnarray}
where $\mathbf{u}_{k}$ is now entirely digital and \ref{eq:f2} is
equivalently written as: 

\begin{eqnarray*}
\mathcal{F}_{main} & : & \underset{\mathbf{Y}_{k}}{\mathrm{\max}\quad}t\\
\mathrm{s.t.} & : & \mathrm{tr}(\mathbf{Q}_{m}\mathbf{Y}_{k})\geq t\\
 &  & \mathrm{tr}(\mathbf{Y}_{k})=N,\\
 &  & \forall k\in\left\{ 1,\ldots,G\right\} ,\forall m\in\left\{ 1,\ldots,M\right\} .
\end{eqnarray*}
where $\mathbf{Y}_{k}=\mathrm{tr}(\mathbf{u}_{k}\mathbf{u}_{k}^{H})$
and $\mathbf{Q}_{m}=\mathrm{tr}(\mathbf{h_{i}}\mathbf{h}_{i}^{H})$.
Problem $\mathcal{F}_{main}$ represents a set of single group max-min
fairness multicasting problems where the problem is always feasible
\cite{Sidiropoulos2006}. In addition, if the solution is optimal,
$\mathcal{F}_{main}$ outputs a rank-one solution. In fact, Problem
$\mathcal{F}_{main}$ always outputs a rank-one solution since the
channels are all rank-one. Hence, the principle component vectors
of $\left\{ \mathbf{Y}_{k}\right\} _{k=1}^{G}$ gives the optimal
digital precoders $\left\{ \mathbf{u}_{k}^{\star}\right\} _{k=1}^{G}$.
Once $\left\{ \mathbf{u}_{k}^{\star}\right\} _{k=1}^{G}$ is obtained
and denoted $\mathbf{U}=[\mathbf{u}_{1}^{\star},\ldots,\mathbf{u}_{N_{RF}}^{\star}]$
, the RF-precoder has a closed-form solution
\begin{equation}
\arg(\mathbf{F})=\arg(\mathbf{U})\label{eq:F_RF}
\end{equation}

After finding the RF precoder based on the criterion described above,
the equivalent channels $\left\{ \mathbf{Q}_{\mathbf{eff},m}=\mathbf{h}_{\mathbf{eff},m}\mathbf{h}_{\mathbf{eff},m}^{H}\right\} _{m=1}^{m=M}$
are formed. Then, Problem $\mathcal{P}_{main}$ is solved, if feasible,
to obtain an upper bound for the max-min SINR value.

\section{SIMULATION RESULTS} \label{numresults}

In this section, simulation results are presented to show the performance
of the hybrid proposed algorithm versus a fully-digital system algorithm
with ($N_{{\scriptscriptstyle \mathrm{RF}}}=N$) for various multicasting
systems. A mmWave channel with $L=1$ is considered between the $m$th
user and the BS and an Uniformally-Linear-Array (ULA) with $N$ elements
is assumed at the BS. The angle of departure (AoD) of the $m$th user,
$\phi_{m}$, is drawn from a uniform random variable defined as $\phi_{m}\sim U(0,2\pi)$.
In addition the channel gain of the $m$th user is drawn from a circularly
symmetric Gaussian random variable defined as $\beta_{m}\sim\mathcal{CN}(0,1)$.
The noise variance of the receivers is assumed as $\sigma_{1}^{2}=\sigma_{2}^{2}=\cdots=\sigma_{M}^{2}=\sigma_{n}^{2}=1$.
The transmit power at the BS is varied from $-10\mathrm{dB}$ to $50\mathrm{dB}$
with $5\mathrm{dB}$ increment. 

\begin{figure}
\begin{centering}
\includegraphics[width = 0.5\textwidth]{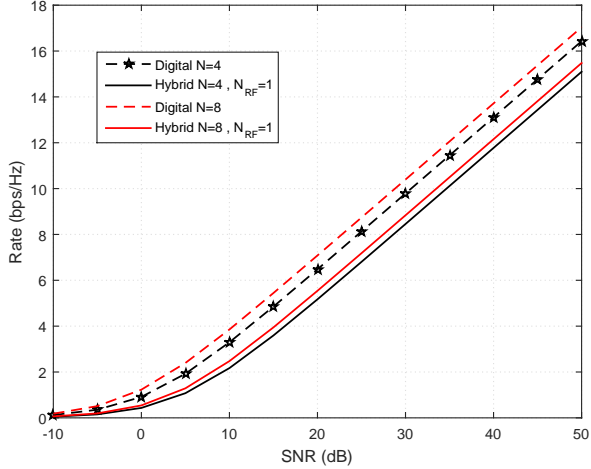}
\par\end{centering}
\caption{Max-min rate of a Single group multicasting system with $G=1$, $M=4$
users. \label{fig:Single-group-multicasting}}
\end{figure}

Fig. \ref{fig:Single-group-multicasting} shows the performance of
the hybrid proposed algorithm versus the fully-digital system algorithm
for a single group multicasting system (broadcasting). In this scenario,
number of users in the group is fixed as $M=4$ while number of antennas
$N$ is varied to evaluate the performance. As can be seen in Fig.
\ref{fig:Single-group-multicasting}, both systems with different
$N$ values have the same slope of $1$ at high SNR values which indicates
an interference-free stream experienced by all users in the system.
The proposed algorithm performance is close to the fully-digital system
at low SNR values and with constant gap at high SNR values. As $N$
increases, the rate increases for both systems which is a result of
the array gain offered by the MISO channel. 

Fig. \ref{fig:Multi-user-beamforming,-} shows the performance of
the hybrid proposed algorithm versus the fully-digital system for
a multi-user beamforming system with $N=64$, $G=3$ and $M=3$. As
can be seen in Fig. \ref{fig:Multi-user-beamforming,-}, the performance
of the proposed algorithm is close to the fully-digital system algorithm
at low SNR values. However, the gap increases as the transmit power
increases. At high SNR values, the interference between different
users becomes more dominant compared to the noise level and thus degrades
the overall performance of the proposed algorithm. The interference
causes the slope of the rate curve to gradually decrease as SNR value
increases until it saturates. Conversely, at low SNR values, the interference
is neglected compared to the noise level and hence the performance
of both algorithms are relatively similar. 

\begin{figure}
\begin{centering}
\includegraphics[width = 0.5\textwidth]{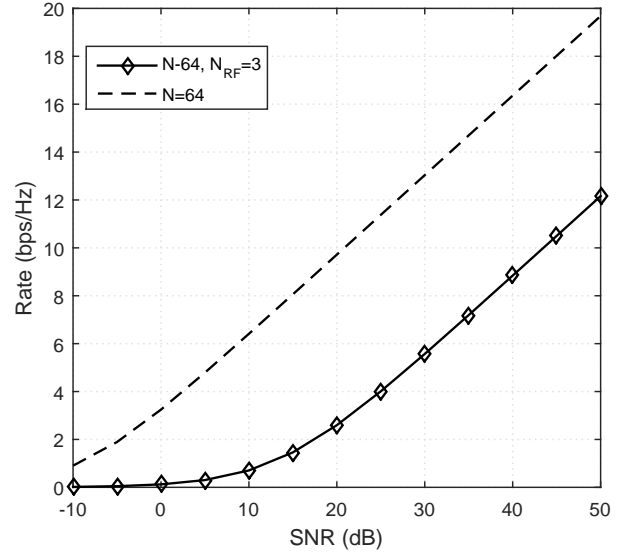}
\par\end{centering}
\caption{Max-min rate of a Multi-user beamforming system with $N=64$ antennas,
$G=3$ groups, $M=3$ users.\label{fig:Multi-user-beamforming,-}}
\end{figure}

Fig. \ref{fig:-antennas,-} shows the performance of both systems
for a multi-group multicasting system with $G=3$, $N_{{\scriptscriptstyle \mathrm{RF}}}=3,$
and $M=6$. Users are equally distributed among groups, i.e. $2$
users per group. Number of antennas $N$ is varied to evaluate the
performance of both systems. As seen in Fig. \ref{fig:-antennas,-},
at low SNR values the performance of both algorithms are extremely
close. However, at high SNR, the rate curves for the proposed algorithm
with different $N$ values saturates while the fully-digital system
experiences no saturation effect. The saturation is due to inter-group
interference experienced by users belonging to different groups. The
fully-digital system doesn't suffer from the interference due to the
minimal adequate amount of antennas at the transmitter to fully null-out
the interference. In specific, for equal user-group distribution,
the minimum number of antennas at the transmitter to null-out the
inter-group interference is $1+M-\mathcal{\left|G\right|}$. Figure
\ref{fig:-antennas,--1} illustrates the performance for an asymmetric
distribution specified in the figure caption compared to the symmetric
distribution (equal user-group distribution). As can be seen in Figure
\ref{fig:-antennas,--1}, with an asymmetric distribution, the rate
curve slope is larger compared to the symmetric case. Saturation occurs
in both cases but it would happen later in the asymmetric scenario
and the reason could attribute to the less interfering users in the
groups with smaller number of users compared to the group with largest
number of users. 

\begin{figure}
\begin{centering}
\includegraphics[width = 0.5\textwidth]{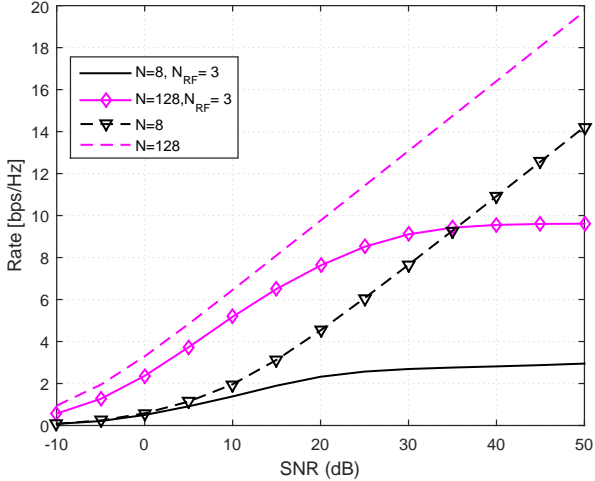}
\par\end{centering}

\caption{Max-min rate of a MGM system with $G=3$ groups, $M=6$ users (equal
group distribution).\label{fig:-antennas,-}}

\end{figure}

\begin{figure}
\centering{}\includegraphics[width = 0.5\textwidth]{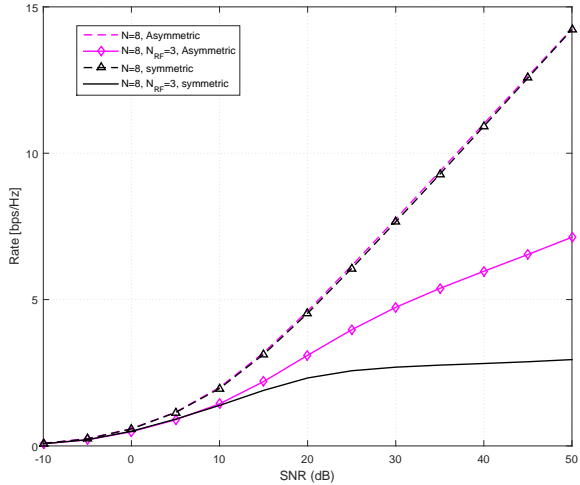}\caption{Max-min rate of a MGM system with $N=8$ antennas, $G=3$ groups,
$M=6$ users ($\mathcal{\left|G\right|}_{1}=3$, $\mathcal{\left|G\right|}_{2}=2$,
$\mathcal{\left|G\right|}_{3}=1$).\label{fig:-antennas,--1}}
\end{figure}

In Fig. \ref{fig:-antennas,--2}, the performance of the two systems
are shown for a multi-group multicasting system with $N=8$, $G=3$,
$M=6$ and different number of paths per user's channel. The goal
is to evaluate the performance at an extreme mmWave case when $L=1$
and a moderate case when $L=15$. For the fully-digital system, the
extra paths in the channel allows for an increase in the rate for
all users. In contrast, for the proposed system design and at high
SNR values, the extra paths result in an early saturation effect compared
to the single path channel performance. Because the design only accounts
for the dominant path of each user, increasing number of paths per
user creates a greater inter-group interference and eventually results
in a saturation effect. At low SNR values, the interference is minimal
and a larger rate is observed when $L=15$ compared to the performance
when $L=1$. 

\begin{figure}
\begin{centering}
\includegraphics[width = 0.5\textwidth]{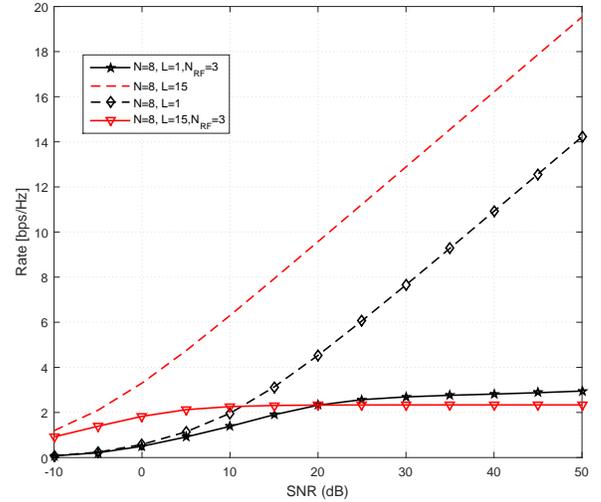}
\par\end{centering}
\caption{Max-min rate of a MGM system with $N=8$ antennas, $G=3$ groups,
$M=6$ users (equal group distribution).\label{fig:-antennas,--2}}
\end{figure}

\section{CONCLUSION} \label{conclusion}
In this paper, a low complexity max-min fair hybrid precoder design was proposed for multi-group multicasting systems. The design extended to multi-user systems as well as broadcasting systems. The design was motivated by the special structure of the channel, namely, single-path mmWave channel for each user. The RF precoder consisted of $G$ analog beamformer(s). Each analog beamformer was designed to maximize the minimum matching component for users within a given group. Once obtained, the digital precoder was attained by solving the max-min problem of the equivalent channel. The performance of the fully-digital system and the proposed system design were shown by simulations for multi-group multicasting, broadcasting and multi-user systems.

\bibliographystyle{IEEEtran}
\bibliography{reference}

\begin{thebibliography}{10}
\providecommand{\url}[1]{#1}
\csname url@samestyle\endcsname
\providecommand{\newblock}{\relax}
\providecommand{\bibinfo}[2]{#2}
\providecommand{\BIBentrySTDinterwordspacing}{\spaceskip=0pt\relax}
\providecommand{\BIBentryALTinterwordstretchfactor}{4}
\providecommand{\BIBentryALTinterwordspacing}{\spaceskip=\fontdimen2\font plus
\BIBentryALTinterwordstretchfactor\fontdimen3\font minus
  \fontdimen4\font\relax}
\providecommand{\BIBforeignlanguage}[2]{{%
\expandafter\ifx\csname l@#1\endcsname\relax
\typeout{** WARNING: IEEEtran.bst: No hyphenation pattern has been}%
\typeout{** loaded for the language `#1'. Using the pattern for}%
\typeout{** the default language instead.}%
\else
\language=\csname l@#1\endcsname
\fi
#2}}
\providecommand{\BIBdecl}{\relax}
\BIBdecl

\bibitem{Silva2009}
Y.~Silva and A.~Klein, ``{Linear Transmit Beamforming Techniques for the
  Multigroup Multicast Scenario},'' \emph{IEEE Transactions on Vehicular
  Technology}, vol.~58, no.~8, pp. 4353--4367, 2009.

\bibitem{Ogunbekun2003}
J.~Ogunbekun, ``{MBMS service provision and its challenges},'' in \emph{Fourth
  International Conference on 3G Mobile Communication Technologies}, vol. 2003,
  no.~I, 2003, pp. 128--133.

\bibitem{Bakhuizen2005}
M.~Bakhuizen and U.~Horn, ``{Mobile broadcast / multicast in mobile
  networks},'' \emph{Ericsson Rev}, no. 01/2005, pp. 1--8, 2005.

\bibitem{Jiang2007}
T.~Jiang, W.~Xiang, H.~H. Chen, and Q.~Ni, ``{Multicast broadcast services
  support in OFDMA-based WiMAX systems},'' \emph{IEEE Communications Magazine},
  vol.~45, no.~8, pp. 78--86, 2007.

\bibitem{Bengtsson1999}
M.~Bengtsson and B.~Ottersten, ``{Optimal Downlink Beamforming Using
  Semidefinite Optimization},'' in \emph{Proc. of 37th Annual Allerton
  Conference}, 1999, pp. 987--996.

\bibitem{MBengtsson2002}
B.~O. {M Bengtsson}, ``{Optimal and Suboptimal Transmit Beamforming},'' in
  \emph{Handbook of Antennas in Wireless Communications}.\hskip 1em plus 0.5em
  minus 0.4em\relax Boca Raton, FL: CRC, 2002, ch.~18, pp. 18--1.

\bibitem{Visotsky1999}
E.~Visotsky and U.~Madhow, ``{Optimum beamforming using transmit antenna
  arrays},'' in \emph{1999 IEEE 49th Vehicular Technology Conference (Cat.
  No.99CH36363)}, vol.~1, no.~2, 1999, pp. 851--856.

\bibitem{Rashid-Farrokhi1998}
F.~Rashid-Farrokhi, K.~Liu, and L.~Tassiulas, ``{Transmit beamforming and power
  control for cellular wireless systems},'' \emph{IEEE Journal on Selected
  Areas in Communications}, vol.~16, no.~8, pp. 1437--1450, 1998.

\bibitem{Yu2007}
W.~Yu and T.~Lan, ``{Transmitter Optimization for the Multi-Antenna Downlink
  With Per-Antenna Power Constraints},'' \emph{IEEE Transactions on Signal
  Processing}, vol.~55, no.~6, pp. 2646--2660, 2007.

\bibitem{Schubert2004}
M.~Schubert and H.~Boche, ``{Solution of the Multiuser Downlink Beamforming
  Problem With Individual SINR Constraints},'' \emph{IEEE Transactions on
  Vehicular Technology}, vol.~53, no.~1, pp. 18--28, 2004.

\bibitem{Sidiropoulos2006}
N.~Sidiropoulos, T.~Davidson, and {Zhi-Quan Luo}, ``{Transmit beamforming for
  physical-layer multicasting},'' \emph{IEEE Transactions on Signal
  Processing}, vol.~54, no.~6, pp. 2239--2251, 2006.

\bibitem{Karipidis2008}
E.~Karipidis, N.~Sidiropoulos, and {Zhi-Quan Luo}, ``{Quality of Service and
  Max-Min Fair Transmit Beamforming to Multiple Cochannel Multicast Groups},''
  \emph{IEEE Transactions on Signal Processing}, vol.~56, no.~3, pp.
  1268--1279, 2008.

\bibitem{Lu2014}
L.~Lu, G.~Y. Li, A.~L. Swindlehurst, A.~Ashikhmin, and R.~Zhang, ``{An Overview
  of Massive MIMO: Benefits and Challenges},'' \emph{IEEE Journal of Selected
  Topics in Signal Processing}, vol.~8, no.~5, pp. 742--758, 2014.

\bibitem{Heath2016}
R.~W. Heath, N.~Gonzalez-Prelcic, S.~Rangan, W.~Roh, and A.~M. Sayeed, ``{An
  Overview of Signal Processing Techniques for Millimeter Wave MIMO Systems},''
  \emph{IEEE Journal of Selected Topics in Signal Processing}, vol.~10, no.~3,
  pp. 436--453, 2016.

\bibitem{Rappaport2013}
T.~S. Rappaport, {Shu Sun}, R.~Mayzus, {Hang Zhao}, Y.~Azar, K.~Wang, G.~N.
  Wong, J.~K. Schulz, M.~Samimi, and F.~Gutierrez, ``{Millimeter Wave Mobile
  Communications for 5G Cellular: It Will Work!}'' \emph{IEEE Access}, vol.~1,
  pp. 335--349, 2013.

\bibitem{Doan2004}
C.~Doan, S.~Emami, D.~Sobel, A.~Niknejad, and R.~Brodersen, ``{Design
  considerations for 60 GHz CMOS radios},'' \emph{IEEE Communications
  Magazine}, vol.~42, no.~12, pp. 132--140, 2004.

\bibitem{Pi2011}
Z.~Pi and F.~Khan, ``{An introduction to millimeter-wave mobile broadband
  systems},'' \emph{IEEE Communications Magazine}, vol.~49, no.~6, pp.
  101--107, 2011.

\bibitem{Zhang2005}
X.~Zhang, A.~F. Molisch, and S.-y. Kung, ``{Variable-phase-shift-based
  RF-baseband codesign for MIMO antenna selection},'' \emph{IEEE Transactions
  on Signal Processing}, vol.~53, no.~11, pp. 4091--4103, 2005.

\bibitem{Alkhateeb2014}
A.~Alkhateeb, {Jianhua Mo}, N.~Gonzalez-Prelcic, and R.~W. Heath, ``{MIMO
  Precoding and Combining Solutions for Millimeter-Wave Systems},'' \emph{IEEE
  Communications Magazine}, vol.~52, no.~12, pp. 122--131, 2014.

\bibitem{Zhang2014}
E.~Zhang and C.~Huang, ``{On achieving optimal rate of digital precoder by
  RF-baseband codesign for MIMO systems},'' in \emph{IEEE Vehicular Technology
  Conference}, Vancouver, BC, 2014, pp. 1 -- 5.

\bibitem{Bogale2016}
T.~E. Bogale, L.~B. Le, A.~Haghighat, and L.~Vandendorpe, ``{On the Number of
  RF Chains and Phase Shifters, and Scheduling Design With Hybrid Analog
  Digital Beamforming},'' \emph{IEEE Transactions on Wireless Communications},
  vol.~15, no.~5, pp. 3311--3326, 2016.

\bibitem{Chae2015}
S.~H. Chae and C.~Jeong, ``{Degrees of Freedom of Interference Channels with
  Hybrid Beam-forming},'' \emph{arXiv:1504.06743 [cs, math]}, pp. 1--27, 2015.

\bibitem{Ayach2014}
O.~E. Ayach, S.~Rajagopal, S.~Abu-Surra, Z.~Pi, and R.~W. Heath, ``{Spatially
  Sparse Precoding in Millimeter Wave MIMO Systems},'' \emph{IEEE Transactions
  on Wireless Communications}, vol.~13, no.~3, pp. 1499--1513, 2014.

\bibitem{Ayach2012}
O.~E. Ayach, R.~W. Heath, S.~Abu-Surra, S.~Rajagopal, and Z.~Pi, ``{Low
  complexity precoding for large millimeter wave MIMO systems},'' in \emph{2012
  IEEE International Conference on Communications (ICC)}, 2012, pp. 3724--3729.

\bibitem{Alkhateeb2015}
A.~Alkhateeb, G.~Leus, and R.~W. Heath, ``{Limited Feedback Hybrid Precoding
  for Multi-User Millimeter Wave Systems},'' \emph{IEEE Transactions on
  Wireless Communications}, vol.~14, no.~11, pp. 6481--6494, 2015.

\bibitem{Sohrabi2015}
F.~Sohrabi and W.~Yu, ``{Hybrid digital and analog beamforming design for
  large-scale MIMO systems},'' in \emph{2015 IEEE International Conference on
  Acoustics, Speech and Signal Processing (ICASSP)}, 2015, pp. 2929--2933.

\bibitem{Alkhateeb2013}
a.~Alkhateeb, O.~{El Ayach}, G.~Leus, and R.~W. Heath, ``{Hybrid precoding for
  millimeter wave cellular systems with partial channel knowledge},'' in
  \emph{2013 Information Theory and Applications Workshop (ITA)}, no.~i, 2013,
  pp. 1--5.

\bibitem{Yu2015}
X.~Yu, J.-c. Shen, J.~Zhang, and K.~B. Letaief, ``{Alternating Minimization
  Algorithms for Hybrid Precoding in Millimeter Wave MIMO Systems},''
  \emph{arXiv:1601.07340v1}, 2016.

\bibitem{Liang2014}
L.~Liang, W.~Xu, and X.~Dong, ``{Low-Complexity Hybrid Precoding in Massive
  Multiuser MIMO Systems},'' \emph{IEEE Wireless Communications Letters},
  vol.~3, no.~6, pp. 653--656, 2014.

\bibitem{Lopez2002}
M.~J. Lopez, ``{Multiplexing, Scheduling, and Multicasting Strategies for
  Antenna Arrays in Wireless Networks},'' Ph.D. dissertation, Massachusetts
  Institute of Technology, 2002.

\bibitem{Luo2010}
Z.-q. Luo, W.-k. Ma, A.~So, Y.~Ye, and S.~Zhang, ``{Semidefinite Relaxation of
  Quadratic Optimization Problems},'' \emph{IEEE Signal Processing Magazine},
  vol.~27, no.~3, pp. 20--34, 2010.

\bibitem{Dai2016}
M.~Dai and B.~Clerckx, ``{Hybrid Precoding for Physical Layer Multicasting},''
  \emph{arXiv:1511.06715v1}, vol.~20, no.~2, pp. 228--231, 2015.

\bibitem{Karipidis2007}
E.~Karipidis, N.~D. Sidiropoulos, and Z.-q. Luo, ``{Far-Field Multicast
  Beamforming for Uniform Linear Antenna Arrays},'' \emph{IEEE Transactions on
  Signal Processing}, vol.~55, no.~10, pp. 4916--4927, 2007.

\bibitem{Christopoulos2014a}
D.~Christopoulos, S.~Chatzinotas, and B.~Ottersten, ``{Sum rate maximizing
  multigroup multicast beamforming under per-antenna power constraints},'' in
  \emph{2014 IEEE Global Communications Conference}, 2014, pp. 3354--3359.

\bibitem{Christopoulos2014b}
------, ``{Multicast multigroup beamforming under per-antenna power
  constraints},'' in \emph{2014 IEEE International Conference on Communications
  (ICC)}, vol.~62, no.~19, 2014, pp. 4704--4710.

\bibitem{Christopoulos2014}
------, ``{Weighted Fair Multicast Multigroup Beamforming Under Per-antenna
  Power Constraints},'' \emph{IEEE Transactions on Signal Processing}, vol.~62,
  no.~19, pp. 5132--5142, 2014.

\bibitem{Universit2011}
N.~Bornhorst and M.~Pesavento, ``{An iterative convex approximation approach
  for transmit beamforming in multi-group multicasting},'' in \emph{2011 IEEE
  12th International Workshop on Signal Processing Advances in Wireless
  Communications}, 2011, pp. 426--430.

\bibitem{Schad2012}
A.~Schad and M.~Pesavento, ``{Max-min fair transmit beamforming for multi-group
  multicasting},'' in \emph{2012 International ITG Workshop on Smart Antennas
  (WSA)}, 2012, pp. 115--118.

\bibitem{Kaliszan2012}
M.~Kaliszan, E.~Pollakis, and S.~Stanczak, ``{Multigroup multicast with
  application-layer coding: Beamforming for maximum weighted sum rate},'' in
  \emph{2012 IEEE Wireless Communications and Networking Conference (WCNC)},
  2012, pp. 2270--2275.

\bibitem{Christopoulos}
D.~Christopoulos, S.~Chatzinotas, and B.~Ottersten, ``{Multicast multigroup
  beamforming for per-antenna power constrained large-scale arrays},'' in
  \emph{2015 IEEE 16th International Workshop on Signal Processing Advances in
  Wireless Communications (SPAWC)}, 2015, pp. 271--275.

\bibitem{Mx1}
M.~Sadeghi, L.~Sanguinetti, and C.~Yuen, ``Hybrid precoding for multi-group
  physical layer multicasting,'' in \emph{European Wireless 2018; 24th European
  Wireless Conference}.\hskip 1em plus 0.5em minus 0.4em\relax VDE, 2018, pp.
  1--6.

\bibitem{Mx2}
M.~Sadeghi, E.~Bj{\"o}rnson, E.~G. Larsson, C.~Yuen, and T.~L. Marzetta,
  ``Max--min fair transmit precoding for multi-group multicasting in massive
  mimo,'' \emph{IEEE Transactions on Wireless Communications}, vol.~17, no.~2,
  pp. 1358--1373, 2017.

\bibitem{QoS1}
J.~Huang, Z.~Cheng, E.~Chen, and M.~Tao, ``Low-complexity hybrid analog/digital
  beamforming for multicast transmission in mmwave systems,'' in \emph{2017
  IEEE International Conference on Communications (ICC)}.\hskip 1em plus 0.5em
  minus 0.4em\relax IEEE, 2017, pp. 1--6.

\bibitem{QoS2}
L.~F. Abanto-Leon, M.~Hollick, and G.~H. Sim, ``Hybrid precoding for
  multi-group multicasting in mmwave systems,'' in \emph{2019 IEEE Global
  Communications Conference (GLOBECOM)}.\hskip 1em plus 0.5em minus 0.4em\relax
  IEEE, 2019, pp. 1--7.

\bibitem{Liang2014a}
L.~Liang, Y.~Dai, W.~Xu, and X.~Dong, ``{How to approach zero-forcing under RF
  chain limitations in large mmWave multiuser systems?}'' in \emph{2014
  IEEE/CIC International Conference on Communications in China (ICCC)}, 2014,
  pp. 518--522.

\end{thebibliography}

\end{document}